\documentclass[aps,prd,twocolumn,superscriptaddress,a4paper]{revtex4}

\usepackage{graphicx}
\usepackage{eurosym}

\begin{document}


\title{First Results of the Phase II SIMPLE Dark Matter Search}


\author{M. Felizardo}
 \affiliation{Department of Physics, Universidade Nova de Lisboa,
 2829-516 Caparica, Portugal} \affiliation{Instituto Tecnol\'ogico e
 Nuclear, Estrada Nacional 10, 2686-953 Sacav\'em, Portugal}
 \affiliation{Centro de F\'isica Nuclear, Universidade de Lisboa,
 1649--003 Lisbon, Portugal}
\author{T. Morlat} \affiliation{Centro de F\'isica Nuclear, Universidade de Lisboa,
1649--003 Lisbon, Portugal} \affiliation {Department of Physics,
Universidade de Lisboa, Campo Grande C8, 1749-016 Lisboa, Portugal }
\author{A.C. Fernandes} \affiliation{Instituto
Tecnol\'ogico e Nuclear, Estrada Nacional 10, 2686-953 Sacav\'em,
Portugal} \affiliation{Centro de F\'isica Nuclear, Universidade de
Lisboa, 1649--003 Lisbon, Portugal}
\author{TA Girard} \email[corresponding  author:]{criodets@cii.fc.ul.pt}
\affiliation{Centro de F\'isica Nuclear, Universidade de Lisboa,
1649--003 Lisbon, Portugal} \affiliation {Department of Physics,
Universidade de Lisboa, Campo Grande C8, 1749-016 Lisboa, Portugal }
\author{J.G. Marques}
\affiliation{Instituto Tecnol\'ogico e Nuclear, Estrada Nacional 10,
2686-953 Sacav\'em, Portugal} \affiliation{Centro de F\'isica
Nuclear, Universidade de Lisboa, 1649--003 Lisbon,
Portugal}
\author{A.R. Ramos} \affiliation{Instituto Tecnol\'ogico e
Nuclear, Estrada Nacional 10, 2686-953 Sacav\'em, Portugal}
\affiliation{Centro de F\'isica Nuclear, Universidade de Lisboa,
1649--003 Lisbon, Portugal}
\author{M. Auguste} \affiliation{Laboratoire Souterrain \`a Bas Bruit, Observatoire de
la C\^ote d'Azur, 84400 Rustrel--Pays d'Apt, France}
\author{D. Boyer} \affiliation{Laboratoire Souterrain \`a Bas Bruit, Observatoire de
la C\^ote d'Azur, 84400 Rustrel--Pays d'Apt, France}
\author{A. Cavaillou} \affiliation{Laboratoire Souterrain \`a Bas Bruit, Observatoire de
la C\^ote d'Azur, 84400 Rustrel--Pays d'Apt, France}
\author{C. Sudre} \affiliation{Laboratoire Souterrain \`a Bas Bruit, Observatoire de
la C\^ote d'Azur, 84400 Rustrel--Pays d'Apt, France}
\author{J. Poupeney} \affiliation{Laboratoire Souterrain \`a Bas Bruit, Observatoire de
la C\^ote d'Azur, 84400 Rustrel--Pays d'Apt, France}
\author{R.F. Payne} \affiliation{Pacific Northwest National Laboratory, Richland, WA
99352 USA}
\author{H.S. Miley} \affiliation{Pacific Northwest
National Laboratory, Richland, WA 99352 USA}
\author{J. Puibasset} \affiliation{Centre de Recherche sur la Mati\'ere Divis\'ee
CNRS et Universit\'e d'Orl\'eans, 45071 Orl\'eans, cedex 02 France }
\collaboration{The SIMPLE collaboration} \noaffiliation

\date{\today}

\begin{abstract}
We report results of a 14.1 kgd measurement with 15 superheated
droplet detectors of total active mass 0.208 kg, comprising the
first stage of a 30 kgd Phase II experiment. In combination with the
results of the neutron-spin sensitive XENON10 experiment, these
results yield a limit of $|a_{p}| < $ 0.32 for M$_{W}$ = 50
GeV/c$^{2}$ on the spin-dependent sector of weakly interacting
massive particle-nucleus interactions with a 50\% reduction in the
previously allowed region of the phase space formerly defined by
XENON, KIMS and PICASSO. In the spin-independent sector, a limit of
2.3$\times$10$^{-5}$ pb at M$_{W}$ = 45 GeV/c$^{2}$ is obtained.

\end{abstract}


\maketitle


The direct search for weakly interacting massive particle (WIMP)
dark matter continues to be among the forefront endeavors of modern
physics activity. Search experiments are generically based on the
detection of nuclear recoil events resulting from WIMP-nucleus
interactions, and are traditionally classified as spin-independent
(SI) or spin-dependent (SD) according to which interaction channel
the experiment is most sensitive, of which the first has generally
attracted the most attention. The current status of the SI search
for WIMPs is defined by a number of projects, including XENON
\cite{xenon}, CDMS \cite{cdms} and ZEPLIN \cite{zeplin}, which as a
result of their target nuclei spins also define the WIMP-neutron
sector of the SD phase space. The WIMP-proton sector is currently
constrained by PICASSO \cite{picasso} and KIMS \cite{kims}.

The SIMPLE (Superheated Instrument for Massive ParticLe Experiments)
\cite{morlat,simple0} project, located in a 61 m$^{3}$ cavern at the
1500 mwe level of the Laboratoire Souterrain \`a Bas Bruit (LSBB) in
southern France, currently runs superheated droplet detectors
(SDDs). The SDD is a suspension of 1-2\% superheated liquid
C$_{2}$ClF$_{5}$ droplets ($\sim$ 30 $\mu$m radius) in a
viscoelastic 900 ml gel matrix which undergo transitions to the gas
phase upon energy deposition by incident radiation. Two conditions
are required for the nucleation of the gas phase in the superheated
liquid \cite{seitz}: (i) the energy deposited must be greater than a
thermodynamically-defined minimum energy, and (ii) this energy must
be deposited within a thermodynamically-defined maximum distance
inside the droplet. Together, energy depositions of order $\sim$ 150
keV/$\mu$m are required for a bubble nucleation, which renders the
SDD effectively insensitive to the majority of traditional detector
backgrounds which complicate more conventional dark matter search
detectors (including electrons, $\gamma$'s and cosmic muons). The
insensitivity is not trivial, comprising an intrinsic rejection
factor superior to that of other search techniques by 1-5 orders of
magnitude. Additional advantages of the superheated technique
include low cost, scalability, and increased sensitivity to
WIMP-proton spin interactions via the $^{19}$F content \cite{mi}.

The SDDs were fabricated according to previously-described
procedures \cite{morlat}, in an underground (210 mwe) clean room in
close proximity to the measurement site. The SIMPLE gel ingredients,
all biologically-clean food products, are purified using
actinide-specific ion-exchanging resins. The freon is single
distilled; the water, double distilled. The presence of U/Th
contaminations in the gel, measured at $\sim$ 0.1 ppb by low-level
$\alpha$ and $\gamma$ spectroscopy of the production gel, yields an
overall $\alpha$-background level of $<$ 0.5 evt/kg freon/d. A
similar level is measured for the detector containment materials.

The detectors are capped using a mechanical construction which
virtually eliminates pressure microleaks \cite{simple0}. Each cap
contains feedthroughs for pressure monitoring, and for a 20-16k Hz
electret microphone encased in a latex sheath, which is immersed in
a 4 cm thick glycerin layer covering the gel at the top of the
detector containment. Each microphone response is preamplified and
recorded in a MatLab platform in sequential files of 8 min duration
\cite{felizardo}, with resolutions of 0.3 mV in amplitude and 1.6
$\times$ 10$^{-2}$ ms in time; the pressure reading is similarly
recorded separately. The use of shielded telecommunications-grade
cabling eliminates signal resulting from cable motion and
parasitics, even when exaggerated.

The SDDs are immersed to a depth of 20 cm in a 700 liter water pool
maintained at a bath temperature of 9.0$\pm 0.1$ $^{o}$C within the
cavern, and pressurized to 2 bar to reduce background sensitivity.
The water pool rests on a dual vibration absorber placed atop a 20
cm thick wood platform resting on a 50 cm thick concrete floor. The
pool is surrounded by three layers of sound and thermal insulation.
An additional 50-75 cm thick water shielding surrounds the insulated
pool and pedestal, with a 75 cm water thickness overhead; 50 cm of
water separates the pool bottom from the SDD bases.

At 1500 mwe, the ambient neutron flux is primarily due to the
surrounding calcite rock, estimated at well-below 4 $\times$
10$^{-5}$ n/cm$^{2}$s \cite{waysand}. The cavern is shielded from
the rock environment by a 30-100 cm thickness of concrete,
internally sheathed by a 1 cm thickness of iron. Radio-assays of the
concrete yielded 1.90$\pm$0.05 ppm $^{232}$Th and 0.850$\pm$0.081
ppm $^{238}$U; of the steel, 3.20$\pm$0.25 ppb $^{232}$Th and
2.9$\pm$0.2 ppb $^{238}$U. The results are at the same level as
those recorded in other underground locations such as Canfranc,
Modane and Gran Sasso \cite{canfranc,modane,gsasso}. Monte-Carlo
simulations of the on-detector neutron field, which include all
shielding materials and account for spontaneous fission plus
decay-induced ($\alpha$,n) reactions, show negligible variations for
concrete thicknesses $\geq$ 20 cm, and yield an expected neutron
background of 1.09 $\pm$ 0.02 (stat) $\pm$ 0.07 (syst) evt/kgd.

The ambient radon level varies seasonally between 28 - 1000
Bq/m$^{3}$ as a result of water circulation in the mountain. The
cavern air is purged $\sim$ 10 times per day, reducing the ambient
radon levels to $\sim$ 60 Bq/m$^{3}$. Diffusion of the environmental
radon into a detector is limited by the surrounding water, which
covered the detectors to just above their glycerin levels, and is
circulated at 25 liter/min (equivalent to replacing the top 1 cm
water layer each minute). The radon contribution is also low because
of the short radon diffusion lengths of the SDD construction
materials (glass, plastic, metal), the N$_{2}$ overpressuring which
inhibits the advective influx of Rn (via stiffening of the gel), and
the glycerin layer covering the gel. The reduction in the overall
radon contribution to the measurement, including its progeny, is
estimated at $\sim$ 10$^{5}$ with the overall $\alpha$ contribution
to the measurement (including the detector contribution) estimated
at 3.26 $\pm$ 0.08 (stat) $\pm$ 0.76 (syst) evt/kgd.

Data obtained from 15 SDDs, containing between 8-21 g of
C$_{2}$ClF$_{5}$ for a total active mass of 0.208 kg, between 27
October 2009 and 05 February 2010 were analyzed for this report. An
additional, similarly installed, freon-less but otherwise identical
SDD, served as an acoustic veto. The total raw exposure was
14.10$\pm$0.01 kgd, with 1.94 kgd resulting from the detectors being
introduced at one device per day over the three week installation
period, and a 4.70 kgd loss from weather-induced power failures
during the run.

The SDD signals, pressures and temperature are monitored
continuously during operation, as also the radon level. Each
detector was first inspected for raw signal rate and pressure
evolution over the measurement period. An initial data set (4056
events) was then formed by passing the data files through a pulse
validation routine \cite{felizardo} which tagged signal events if
their amplitude exceeded the noise level of the detector by 2 mV.
Tagged signals in coincidence with the freon-less device were next
rejected, as also all candidate signals with less than five pulse
spikes above threshold; the remaining set was then cross-correlated
in time between all SDDs, and coincidences rejected as local noise
events and that a WIMP interacts with no more than one of the
in-bath detectors.

\begin{figure}[h]
  \includegraphics[width=8 cm]{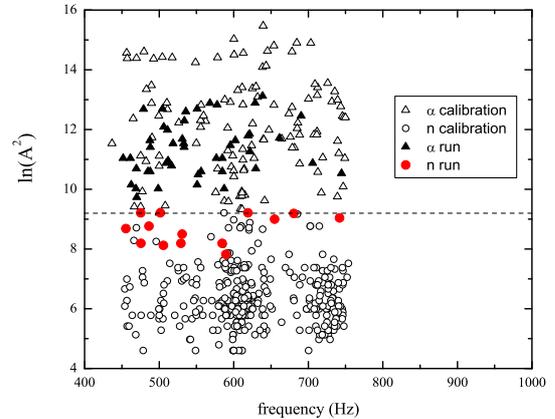}\\
  \caption{Scatter plot of the squared amplitude and frequency
  of the primary harmonic of each true nucleation event (solid), together
  with the same for neutron and $\alpha$ calibration events (open), with
  the dashed horizontal line corresponding to a signal amplitude of 100 mV.}
  \label{fig3}
\end{figure}

The signal waveform, decay time constant and spectral density
structure of the remaining 1828 single events were next inspected
individually. A particle-induced nucleation event possesses a
characteristic frequency response, with a time span of a few
milliseconds, a decay constant of 5-40 ms, and a primary harmonic
between 0.45-0.75 kHz \cite{felizardo}. This response differs
significantly from those of gel-associated acoustic backgrounds such
as trapped N$_{2}$ gas (3.4\% of the total), gas escape (0.008\%),
and gel fractures (4.4\%) which appear at lower frequencies
\cite{felizardo}, as well as local acoustic backgrounds (88\%) such
as water bubbles which differ in power spectra. This event-by-event
analysis permits isolation of the true nucleation events with an
efficiency of better than 97\% at 95\% C.L. Fig. 1 displays the
signal amplitudes (A) with frequency for each of the identified 60
particle-induced signal events, in which a gap corresponding to A =
100-130 mV is discernible.

At 9$^{o}$C, the reduced superheat of the devices is 0.3, and the
probability of events from electrons, $\gamma$'s and mip's
negligible \cite{morlat} over the exposure. Calibrations of the
$\alpha$ response have been made by doping the devices with U308
during fabrication. The event signals, identified in the same
fashion as described above, are shown in Fig. 1. The SRIM-calculated
dE/dx for $\alpha$'s in C$_{2}$ClF$_{5}$ has a Bragg peak which sets
a lower energy threshold at 200 keV for the operating temperature
and pressure; below this threshold, $\alpha$'s are detected only
through $\alpha$-induced nuclear recoils.

Calibration of the high concentration SDD response to neutrons,
using sources of Am/Be, yielded a minimum threshold recoil energy
(E$_{thr}$) of 8.0$\pm$0.1 keV, with an acoustic detection
efficiency of 0.98$\pm$0.03 at 9$^{\circ}$C and 2 bar. The events
are also displayed in Fig. 1, all of which occur with amplitudes
$\leq$ 100 mV. The difference in the two distributions, particularly
at lower amplitudes, results from performing the calibrations with a
15 cm water shield to enhance the tails on the moderated neutron
spectrum.

Fig. 2 displays a typical histogram of both calibration signal
amplitudes. As seen, the neutron-induced events are of lower
amplitude than the $\alpha$-induced, and empirically well fit by a
Gaussian plus constant background from which a discrimination cut
for A $\leq$ 100 mV is placed with an acceptance of $>$ 97\%. The
small droplet size provides a natural lower cutoff to the deposited
$\alpha$ energy and is together with the dE/dx responsible both for
the amplitude gap between the $\alpha$ and neutron populations, and
the spectral asymmetry in the $\alpha$ distribution.

\begin{figure}[h]
  \includegraphics[width=8 cm]{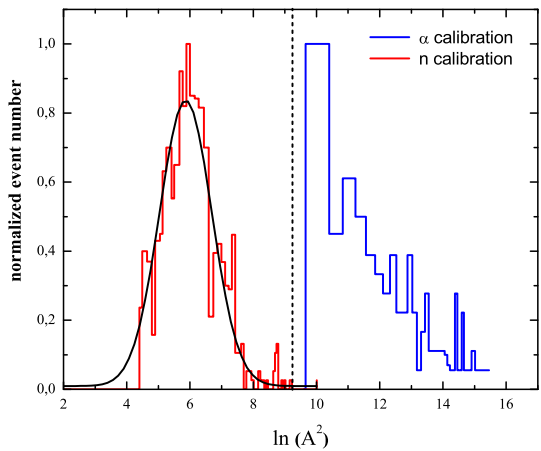}\\
  \caption{A typical histogram of calibration neutron and $\alpha$
  amplitudes, each population consisting of several hundred events.
  The vertical dashed line indicates the neutron discrimination cut
  at A $\leq$ 100 mV, which includes all neutron calibration events.}
  \label{fig3}
\end{figure}

The 14 low amplitude events of the run are consistent with the
neutron calibrations, yielding 0.99 $\pm$ 0.27 (stat) evt/kgd.
Corrected for detection and identification efficiencies, the
difference between measurement and neutron background estimate is
better than 0.3$\sigma$.

An upper limit on the number of WIMP events in the presence of the
uncertain neutron background is estimated by applying the
Feldman-Cousins method \cite{feldman}, based on observing 14 events
with a background one standard deviation below the central value of
the expected neutron background. This yields 4.3 events, and a
resulting WIMP rate of 0.57 evt/kgd at 90\% C.L.

\begin{figure}[h]
  \includegraphics[width=9 cm]{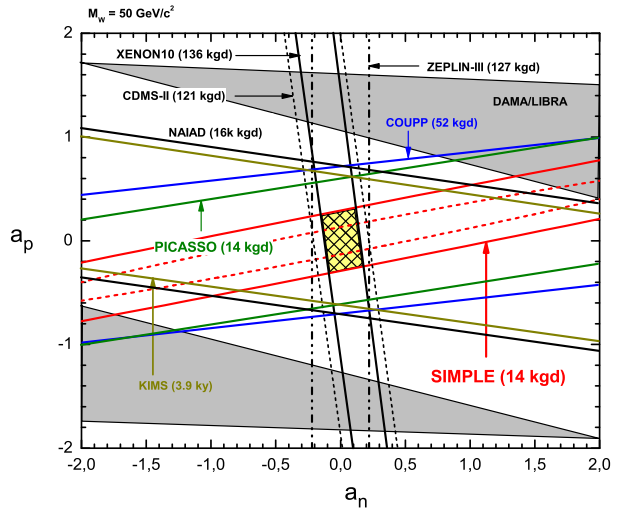}\\
  \caption{a$_{p}$-a$_{n}$ for SIMPLE at M$_{W}$ = 50 GeV/c$^{2}$ ,
together with benchmark experiment results; the dashed contour
represents a "0 evt" SIMPLE result for the same exposure. The
allowed regions are defined by a band (single nuclei target) or
ellipse (multinuclei target), with the external area excluded. The
cross-hatched central area about (0,0) indicates the region allowed
by this result and XENON.} \label{fig4}
\end{figure}

The impact of the result in the SD phase space at M$_{W}$ = 50
GeV/c$^{2}$ is shown in Fig. 3. The contour is calculated within a
model-independent formulation \cite{mi}, in which the region
excluded by an experiment lies outside the indicated band, and the
allowed region is defined by the intersection of the various bands.
The calculations use a standard isothermal halo, bubble nucleation
efficiency of (1-E$_{thr}$/E) \cite{morlat}, and WIMP scattering
rate \cite{lewin} with zero momentum transfer spin-dependent cross
section $\sigma_{SD}$ for elastic scattering:

\begin{equation}\label{vmin}
    \sigma_{SD} \sim  G_{F} [ a_{p}<S_{p}>+ a_{n}<S_{n}> ]^{2}\frac{J+1}{J} ,
\end{equation}

\noindent where $a_{p,n}$ are the WIMP-proton,neutron coupling
strengths, $<S_{p,n}>$ are the expectation values of the proton
(neutron) group's spin, G$_{F}$ is the Fermi coupling constant, and
J is the total nuclear spin. The form factors of Ref. \cite{lewin}
have been used for all odd-A nuclei. The spin values of Strottman
have been used for $^{19}$F \cite{strottman}; for $^{35}$Cl and
$^{37}$Cl, $<S_{p,n}>$ are from Ref. \cite{mi}, while for $^{13}$C
the $<S_{p,n}>$ were estimated by using the odd group approximation.
Use of the Divari et. al. spin values \cite{vergados} for $^{19}$F
would rotate the ellipse about the origin to a more horizontal
position. We include only the 121 kgd result of CDMS \cite{cdms}
since the more recent \cite{newcdms} model-independent result hasn't
yet been published. The shaded area represents the allowed
DAMA/LIBRA region \cite{savage}, which is already excluded by other
experiments.

As indicated, the present result combined with XENON10 yields limits
of $|a_{p}| \leq $ 0.32, $|a_{n}| \leq $ 0.16 on the SD sector of
WIMP-nucleus interactions for M$_{W}$=50 GeV/c$^{2}$, with $\sim$
50\% reduction in the allowed region of the phase space. M$_{W}$
above or below this choice yield less restrictive limits \cite{mi}.

\begin{figure}[h]
  \includegraphics[width=9.5 cm]{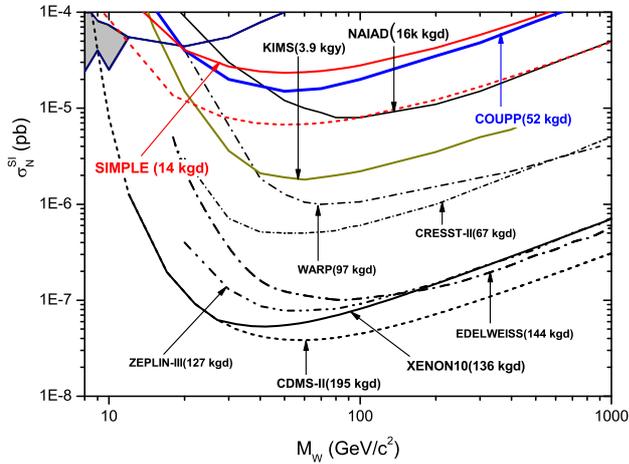}\\
  \caption{Spin-independent contour for SIMPLE; the
  dashed contour represents a "0 evt" SIMPLE result for the same
  measurement exposure. Also shown are several
  \cite{picnote} of the leading spin-independent search results;
  the shaded area represents the recent result of CoGeNT \cite{cogent}. } \label{fig5}
\end{figure}

For completeness, the impact of the result in the SI sector,
calculated following the standard isothermal halo and WIMP elastic
scattering rate of Ref. \cite{lewin} using a Helm form factor, is
shown in Fig. 4 in comparison with results from other leading search
efforts \cite{picnote}. The figure indicates a contour minimum of
2.3$\times$10$^{-5}$ pb at 45 GeV/c$^{2}$, slightly less restrictive
than that of the 52 kgd COUPP \cite{coupp} exposure. The
near-equivalence of the two, despite the $\sim$ 48$\times$
difference in sensitivity from coherence enhancement, most likely
derives from the high radiopurity and $\alpha$-discrimination of the
SIMPLE detectors.

In summary, a conservative analysis of 14.1 kgd of data from the
first phase run of SIMPLE yields new restrictions on the WIMP-proton
parameter space of SD WIMP interactions which improves on those from
both PICASSO and COUPP. These results represent the first stage of
the ongoing 30 kgd Phase II exposure permitted by current funding,
and further demonstrates the competitiveness of the superheated
liquid technique in the search for astroparticle dark matter in both
spin -dependent and -independent sectors. To further approach the
dashed contours of SIMPLE in Figs. 3 and 4 requires further
reduction and/or elimination of the neutron background: improvements
in the second stage of this measurement include a 10 cm increase in
the wood pedestal beneath the waterpool, and installation of an
additional 10 cm polyethylene beneath the detectors, which together
are projected to reduce the overall neutron contribution by $\sim$
factor 5. This is accompanied by additional $\alpha$ and neutron
calibrations, and a refined $\alpha$-neutron discrimination analysis
\cite{russo} towards understanding the differences between these
results and those of PICASSO \cite{aubin}.

We thank Dr. F. Giuliani for numerous suggestions and advices, Dr.
P. Loaiza for the radioassays of the site concrete and steel, Engº
J. Albuquerque of CRIOLAB, Lda for numerous technical assistances
during the measurement staging, and the Casoli's for their
hospitality during our various residences near the LSBB. This work
was supported in part by grant PDTC/FIS/83424/2006 of the Portuguese
Foundation for Science and Technology (FCT), and by the Nuclear
Physics Center of the University of Lisbon.


\begin{thebibliography}{21}
\bibitem[{\citenamefont{pic}(2009)}]{picnote}
\bibinfo{journal}{The PICASSO impact is not indicated since their contour hasn't yet been presented.}
\end{thebibliography}

\end{document}